\documentclass[preprint,aps,amssymb,nofootinbib,showpacs]{revtex4}
\usepackage{graphics}
\usepackage{epsfig}

\def\bfsigma{\mbox{\boldmath $\sigma$}}

\def\lQ{\Lambda_{\rm QCD}}
\newcommand{\nn}{\nonumber}
\newcommand{\be}{\begin{equation}}
\newcommand{\ee}{\end{equation}}
\newcommand{\bea}{\begin{eqnarray}}
\newcommand{\eea}{\end{eqnarray}}
\def\als{\alpha_{\rm s}}
\def\siml{{\ \lower-1.2pt\vbox{\hbox{\rlap{$<$}\lower6pt\vbox{\hbox{$\sim$}}}}\ }}
\def\simg{{\ \lower-1.2pt\vbox{\hbox{\rlap{$>$}\lower6pt\vbox{\hbox{$\sim$}}}}\ }}
\newcommand{\MS}{{\overline{\rm MS}}}

\begin{document}

\preprint{IFUM-810-FT}

\title{The $1P$ quarkonium fine splittings at NLO}
\author {Nora Brambilla and Antonio Vairo}
\affiliation{INFN and Dipartimento di Fisica dell'Universit\`a di Milano \\
via Celoria 16, 20133 Milan, Italy}

\begin{abstract}
We calculate the $1P$ heavy quarkonium fine splittings at NLO
and discuss the impact of the calculation on the $\chi_b(1P)$ splittings.
\end{abstract}


\maketitle

\section{Introduction}
\label{secint}
The calculation of the heavy quarkonium potentials and spectra in perturbation 
theory has a long history \cite{Gupta:1981pd,Buchmuller:1981aj,Pantaleone:1985uf,Titard:1993nn,
Titard:1994id,Titard:1994ry,Pineda:1997hz}. Early calculations 
had to face two main problems: (1) the consistent inclusion of non-perturbative 
effects and (2) the poor convergence of the perturbative series. 
Recently remarkable progress has been made in both.
Non-relativistic effective field theories of QCD, have provided a systematic 
way to factorize non-perturbative effects in heavy quarkonium observables.
Moreover, they have proved useful in organizing and cancelling   
renormalon singularities and in resumming logs with renormalization group techniques:
these being the main sources of large corrections in the perturbative series. 
For a recent review on effective field theories for heavy quarkonium 
we refer to \cite{Brambilla:2004jw}.

This progress has triggered in the last few years a renewed interest 
in perturbative calculations of the heavy quarkonium spectrum. 
In \cite{Brambilla:2001fw,Brambilla:2001qk} the bottomonium spectrum 
was calculated in a fully analytical NNLO calculation that  
implemented the leading order renormalon cancellation.
In \cite{Recksiegel:2002za,Recksiegel:2003fm} a numerical calculation 
of the spectrum was done that included NLO spin-dependent potentials.
This calculation also relies (weakly, according to the authors) 
on some assumption about the long-range behaviour of the static potential. 
In \cite{Kniehl:2003ap,Penin:2004xi} the hyperfine splittings of 
the $1S$ bottomonium, charmonium and $\bar{b}c$ system were calculated 
at NLL accuracy. In general these calculations show rather stable (in the normalization 
scale) and convergent (in the perturbative series) results. 

In this work we calculate the $n=2$, $l=1$ quarkonium fine splittings at NLO.
Some contributions were calculated in \cite{Titard:1993nn}, but 
a complete analytical calculation was missing up to now.
The calculation may be relevant for heavy quarkonium states 
for which the momentum transfer scale $p$ is much larger than 
the scale of non-perturbative physics $\lQ$. 
Two situations may occur under this condition \cite{Brambilla:1999xf}. 
Let us call $E$ the typical kinetic energy of the  heavy quark and antiquark
in the centre-of-mass reference frame: in a non-relativistic bound state $p \gg E$. 
The first situation corresponds to quarkonium states for which $E\simg\lQ$.
Under this circumstance the heavy quarkonium potential is purely perturbative 
and non-perturbative contributions are parametrically suppressed 
with respect to the NLO corrections. 
The second situation corresponds to quarkonium states for which 
$p\gg\lQ\gg E$. Under this circumstance the potential contains a  
perturbative part and short-range non-perturbative contributions. 
A NLO calculation may help in this case to constrain the size of the 
non-perturbative contributions affecting the spin-dependent potentials, which 
have been since long the object of intense study \cite{Schnitzer:1978kf}.
Moreover, due to Poincar\'e invariance, they are related by some exact
relation to the non-perturbative contributions in the static potential \cite{Gromes:1984ma}.

The paper is organized as follows. In Sec.~\ref{secnlo}  we will calculate the  
fine splittings for $n=2$, $l=1$ quarkonium states at NLO. Our main result is Eq. (\ref{1Phfs}).
In Sec.~\ref{secres} we will apply our result to the fine splittings of 
the $\chi_b(1P)$ states.

\section{$n=2$, $l=1$  fine splittings at NLO}
\label{secnlo}
We want to calculate the energy splitting 
\be 
E(1^3P_j) - E(1^3P_{j'}),
\ee
at NLO accuracy. In order to do this we need the relevant spin-dependent potentials and 
the wave function at NLO accuracy.

The spin-orbit, $V_{\rm LS}$, and tensor, $V_{\rm T}$, potentials at NLO can be found, for instance, 
in \cite{Pantaleone:1985uf}; they read:
\bea
V_{\rm LS}({\bf  r}) &=& V_{\rm LS}^{(0)}({\bf  r}) + V_{\rm LS}^{(1)}({\bf  r}),
\label{VLS}
\\
V_{\rm LS}^{(0)}({\bf  r}) &=& {3 C_F \als(\mu) \over 2 m^2 r^3} \,{\bf  L} \cdot {\bf  S},
\label{VLS0}
\\
V_{\rm LS}^{(1)}({\bf  r}) &=& V_{\rm LS}^{(0)}({\bf  r}) \,
{\als\over \pi}\left[
{\beta_0\over 2} (\ln r \, \mu +\gamma_E) - {2\over 3} C_A (\ln m \, r + \gamma_E) 
- {11\over 36}C_A + {2\over 3} C_F + {1\over 9} T_F n_f \right],
\label{VLS1}
\\
\nn
\\
V_{\rm T}({\bf  r}) &=& V_{\rm T}^{(0)}({\bf  r}) + V_{\rm T}^{(1)}({\bf  r}),
\label{VT}
\\
V_{\rm T}^{(0)}({\bf  r}) &=& {C_F \als(\mu)\over 4 m^2 r^3}\, {\bf S}_{12},
\label{VT0}
\\
V_{\rm T}^{(1)}({\bf  r}) &=& V_{\rm T}^{(0)}({\bf  r}) \, 
{\als\over \pi}\left[{\beta_0 \over 2}(\ln r \, \mu +\gamma_E) - C_A (\ln m \, r + \gamma_E)  
+ {1\over 4} C_A + C_F  + {1\over 3} T_F n_f \right],
\label{VT1}
\eea
where ${\bf  L}= {\bf  r} \times {\bf  p}$, ${\bf  S}= {\bf  S}_1 + {\bf  S}_2$, ${\bf S}_i = \bfsigma_i/2$, 
${\bf S}_{12}$ $=$  $3 {\hat {\bf r}}\cdot \bfsigma_1 \,{\hat {\bf r}}\cdot \bfsigma_2$ $-$ $\bfsigma_1\cdot \bfsigma_2$, 
$ C_A = N_c = 3$, $C_F = 4/3$, $T_F=1/2$, $\beta_0 =  11 \, N_c/3 -  4 \, T_F\, n_f/3$, 
$n_f$ is the number of active massless flavours, $\als$ is the strong coupling constant 
in the $\MS$ scheme and $\gamma_E \simeq 0.577216$ is the Euler constant.

In order to calculate the wave function at NLO we have to consider 
the heavy quarkonium Hamiltonian at NLO. In the centre-of-mass reference 
frame it reads \cite{Billoire:1979ih}:
\bea
H &=& {{\bf p}^2 \over m} + V^{(0)}(r) + \delta V^{(0)}(r),
\label{hamnlo}
\\
V^{(0)}(r) &=&  - {C_F \als(\mu) \over r},
\label{v0lo}
\\
\delta V^{(0)}(r) &=& V^{(0)}(r) \, {\als \over \pi} \, 
\left[
{\beta_0 \over 2}(\ln r\,\mu + \gamma_E) 
+ {31\over 36}C_A - {5\over 9} T_F n_f
\right]. 
\label{v0nlo}
\eea
We call $|nljs \rangle$ the eigenstates common to $H$, ${\bf L}^2$, 
${\bf J}^2$ (${\bf J} = {\bf L} + {\bf S}$) and ${\bf S}^2$. At NLO  $|nljs \rangle$ is given by
\bea
|nljs \rangle  &=& |nljs \rangle^{(0)} + |nljs \rangle^{(1)},
\eea
where
\bea
\left( {{\bf p}^2 \over m} + V^{(0)}(r)\right) |nljs \rangle^{(0)} &=& E_n^{(0)}|nljs \rangle^{(0)},
\quad \hbox{with} \quad E_n^{(0)} = - m \left( { C_F\als  \over 2 n} \right)^2, 
\\
|nljs \rangle^{(1)} &=& \sum_{(k\neq n), l'j's'} |k l'j's'\rangle^{(0)}  
{^{(0)}\langle k l'j's' |  \delta V^{(0)}(r) |nljs \rangle^{(0)}  \over E_n^{(0)} - E_k^{(0)}}.
\eea
For later use we notice that \cite{BS}
\bea 
\langle nljs|\, f(r) \, {\bf L}\cdot {\bf S} \, |n'l'j's' \rangle 
&=& \delta_{ll'}\delta_{jj'} \delta_{ss'} {j(j+1)-l(l+1)-s(s+1) \over 2} \langle nljs|\, f(r) \, |n'ljs \rangle ,
\label{YY2}
\\
\langle nljs|\, f(r) \, {\bf S}_{12}  \, |n'lj's' \rangle 
&=& \delta_{jj'}\delta_{ss'}\langle {\bf S}_{12} \rangle_{ljs} \langle nljs|\, f(r) \, |n'ljs \rangle,
\label{YY3}
\eea
where 
\bea
\langle {\bf S}_{12} \rangle_{ljs} &=&
  \left\{
 \begin{array}{ll}
    - {2l+2 \over 2l -1}  &,\, j=l-1 \\
     + 2                  &,\, j= l   \\
     - {2l \over 2l+3}  &, \, j= l+1
 \end{array}
  \right.
  \ \ \  ,\; {\rm for}\; l\neq 0  \; .
\nn
\\
\nn
\\
\langle {\bf S}_{12}\rangle_{lj0} &=& \langle {\bf S}_{12}\rangle_{0js} = 0\,,
\nn
\eea
and that 
\be
{}^{(0)}\langle nljs|\,f(r)\,|n'l'j's'\rangle^{(0)} = 
\delta_{ll'}\delta_{jj'}\delta_{ss'} {}^{(0)}\langle nl|\,f(r)\,|n'l\rangle^{(0)},
\label{nljsnl}
\ee
where $|nl\rangle^{(0)}$ are the eigenstates common to $[{{\bf p}^2 / m} + V^{(0)}(r)]$  and ${\bf L}^2$.

The fine splitting $E(1^3P_j) - E(1^3P_{j'})$ at NLO is given by
\be 
E(1^3P_j) - E(1^3P_{j'}) = 
  \langle 21j1| V_{\rm LS} + V_{\rm T} |21j1\rangle 
- \langle 21j'1| V_{\rm LS} + V_{\rm T} |21j'1\rangle,
\ee
where 
\be 
\langle 21j1| V_{\rm LS,T} |21j1\rangle = {}^{(0)}\langle 21j1| V_{\rm LS,T} |21j1\rangle^{(0)}
+ {}^{(1)}\langle 21j1| V_{\rm LS,T}^{(0)} |21j1\rangle^{(0)}
+ {}^{(0)}\langle 21j1| V_{\rm LS,T}^{(0)} |21j1\rangle^{(1)}.
\ee
Using the explicit expressions of the potentials, Eqs. (\ref{VLS})-(\ref{VT1}), we obtain:
\bea
{}^{(0)}\langle 21j1| V^{(0)}_{\rm LS} |21j1 \rangle^{(0)} &=&
m\, {(C_F \, \als(\mu))^4 \over 256} \left[ j(j+1) - 4\right],
\label{expV0LS}
\\
{}^{(0)}\langle 21j1| V^{(1)}_{\rm LS} |21j1 \rangle^{(0)} &=&
{}^{(0)}\langle 21j1| V^{(0)}_{\rm LS} |21j1 \rangle^{(0)}
\nn\\
&&
\hspace{-35mm}
\times 
{\als\over \pi}  \,
\left[
{\beta_0\over 2} \left( \ln {2 \mu \over m C_F \als} +1 \right)
-{2\over 3} C_A  \ln {2 \over C_F \als} 
-{35\over 36}C_A + {2\over 3}C_F + {1\over 9} T_Fn_f
\right] ,
\label{expV1LS}
\\
\nn
\\
{}^{(0)}\langle 21j1| V^{(0)}_{\rm T} |21j1 \rangle^{(0)} &=&
m\, {(C_F \, \als(\mu))^4 \over 768} \langle {\bf S}_{12} \rangle_{1j1},
\label{expV0T}
\\
{}^{(0)}\langle 21j1| V^{(1)}_{\rm T} |21j1 \rangle^{(0)} &=&
{}^{(0)}\langle 21j1| V^{(0)}_{\rm T} |21j1 \rangle^{(0)}
\nn\\
&&
\hspace{-35mm}
\times 
{\als\over \pi}  \,
\left[
{\beta_0\over 2} \left(\ln {2 \mu \over m C_F \als} +1\right)
- C_A  \ln {2 \over C_F \als} 
-{3\over 4}C_A + C_F + {1\over 3} T_F n_f
\right] .
\label{expV1T}
\eea
Equations (\ref{expV1LS}) and (\ref{expV1T}) correct the analogous 
expressions that may be found in \cite{Titard:1993nn,Titard:1994id}.\footnote{
They should read (for $l \neq 0$):
\begin{eqnarray*}
&& {}^{(0)}\langle nljs| V^{(0)}_{\rm LS} |nljs \rangle^{(0)} =
m\, {3 (C_F \, \als(\mu))^4 \over 16 n^3 l (l+1)(2l+1)} \left[ j(j+1) - l(l+1) -s(s+1)\right]
\\
&& \qquad 
\times \left\{
1+ {\als\over \pi}  \, \left[
{\beta_0\over 2} \left(\ln {n \mu \over m C_F \als} -\psi(n+l+1) +\psi(2l+3) +\psi(2l) +\gamma_E 
- {n-l-1/2 \over n} \right)
\right.
\right.
\\
&& \qquad\qquad  
- {2\over 3} C_A  \left(\ln {n \over  C_F \als} -\psi(n+l+1) +\psi(2l+3) +\psi(2l) +\gamma_E 
- {n-l-1/2 \over n} \right)
\\
&& \qquad\qquad  
\left.\left.
-{11\over 36}C_A + {2\over 3}C_F + {1\over 9} T_F n_f
\right]\right\},
\\
\\
&& {}^{(0)}\langle nljs| V^{(0)}_{\rm T} |nljs \rangle^{(0)} =
m\, {(C_F \, \als(\mu))^4 \over 16 n^3 l (l+1)(2l+1)} \langle {\bf S}_{12} \rangle_{ljs}
\\
&& \qquad 
\times \left\{
1+ {\als\over \pi}  \, \left[
{\beta_0\over 2} \left(\ln {n \mu \over m C_F \als} -\psi(n+l+1) +\psi(2l+3) +\psi(2l) +\gamma_E 
- {n-l-1/2 \over n} \right)
\right.
\right.
\\
&& \qquad\qquad  
- C_A  \left(\ln {n \over  C_F \als} -\psi(n+l+1) +\psi(2l+3) +\psi(2l) +\gamma_E 
- {n-l-1/2 \over n} \right)
\\
&& \qquad\qquad  
\left.\left.
+{1\over 4}C_A + C_F + {1\over 3} T_F n_f
\right]\right\},
\end{eqnarray*}
where $\psi$ is the derivative of the logarithm of the Gamma function.
We thank F.~J.~Yndurain for communications on this point.}

To calculate ${}^{(1)}\langle 21j1| V^{(0)}_{\rm LS,T} |21j1 \rangle^{(0)}$ we use 
Eqs. (\ref{YY2}), (\ref{YY3}), (\ref{nljsnl}) and the momentum-space representation given in \cite{Brambilla:1995mg}:
\bea
&& \sum_{(k\neq 2), l} 
{\langle {\bf p} |k l\rangle^{(0)} \;{}^{(0)}\langle k l| {\bf q} \rangle 
\over  E_2^{(0)} - E_k^{(0)}}
= 
{(2 \pi )^3 \delta^3 ({\bf p} - {\bf q} ) \over  
-\gamma_2^2/m - {\bf p}^2 / m }
\nn\\
&&
\qquad\qquad 
- {1\over -\gamma_2^2/m - {\bf p}^2 / m} \; {4\pi \, C_F \, \als \over |{\bf p}- {\bf q}|^2} \;
  {1\over -\gamma_2^2/m - {\bf q}^2 / m} 
\nn 
\\
&& \qquad\qquad 
- {128 \pi m \gamma_2^3 \over ( {\bf p}^2 + \gamma_2^2)^2 
({\bf q}^2 + \gamma_2^2 )^2} 
\Bigg\{{2 \gamma_2^2 ( {\bf p} - {\bf q} )^2 \over 
({\bf p}^2 + \gamma_2^2) ( {\bf q}^2 + \gamma_2^2 )}
\left( -{9\over 2} + { 6 \gamma_2^2 \over { \bf p}^2 + \gamma_2^2 }
+ {6 \gamma_2^2 \over {\bf q}^2 + \gamma_2^2 } \right)  
\nn
\\
&& \qquad\qquad\qquad\qquad
+ {3 \over 2} - { 4 \gamma_2^2 \over {\bf p}^2 + \gamma_2^2 } 
- { 4 \gamma_2^2\over {\bf q}^2 + \gamma_2^2 } 
\nn
\\
&& \qquad\qquad\qquad\qquad
- \left( {1 \over 2 C_2} - 1\right) \ln C_2 + 
{ 2 C_2 -4 + 1 / C_2 \over \sqrt{4 C_2 -1}}\,  
{\rm arctan} \sqrt{4 C_2-1} \Bigg\}, 
\eea
with
$$
C_2 = { ( {\bf p}^2 + \gamma_2^2 ) ( {\bf q}^2 + \gamma_2^2 ) 
\over 4 \gamma_2^2 ( {\bf p} - {\bf q} ) ^2 }, \quad \hbox{and} \quad 
\gamma_2 =  m \, C_F \als / 4.
$$ 
Eventually we obtain:
\bea
{}^{(1)}\langle 21j1| V^{(0)}_{\rm LS} |21j1 \rangle^{(0)}  &=& 
{}^{(0)}\langle 21j1| V^{(0)}_{\rm LS} |21j1 \rangle^{(1)}  = 
{}^{(0)}\langle 21j1| V^{(0)}_{\rm LS} |21j1 \rangle^{(0)}
\nn\\
&& \times {\als \over \pi}
\left[ {3\over 4} \beta_0 \left( {155\over 54} - {2\over 9}\pi^2 + \ln {2 \mu  \over m C_F \als} \right)
+{31\over 24}C_A - {5\over 6}T_Fn_f \right],
\label{expVLS10}
\\
\nn
\\
{}^{(1)}\langle 21j1| V^{(0)}_{\rm T} |21j1 \rangle^{(0)} &=& 
{}^{(0)}\langle 21j1| V^{(0)}_{\rm T} |21j1 \rangle^{(1)} = 
{}^{(0)}\langle 21j1| V^{(0)}_{\rm T} |21j1 \rangle^{(0)}
\nn\\
&& \times 
{\als \over \pi}
\left[
{3\over 4} \beta_0 \left( {155\over 54} - {2\over 9}\pi^2 + \ln {2 \mu  \over m C_F \als} \right)
+ {31\over 24}C_A - {5\over 6}T_Fn_f \right].
\label{expVT10}
\eea
In Eqs. (\ref{expVLS10}) and (\ref{expVT10}) the logs agree with what calculated in \cite{Titard:1993nn} 
using a variational method and the last two terms proportional to $C_A$ and $T_Fn_f$ may be also derived 
from Eqs. (\ref{v0nlo}), (\ref{expV0LS}) and (\ref{expV0T}). The other terms are the new contributions to the fine 
splittings at NLO provided by this work. 

Summing up all contributions, the fine splittings of the $n=2$, $l=1$ quarkonium states 
are given at NLO accuracy by
\bea
&&E(1^3P_j) - E(1^3P_{j'}) = { m (C_F \, \als)^4 \over 768}
\nn\\
&&
\times \Bigg\{ 3 \left[ j(j+1) - j'(j'+1) \right]  
\Bigg( 1+ {\als \over \pi}
\left[ \beta_0 \left( 2 \ln {2 \mu  \over m C_F \als} + {173\over 36} - {\pi^2 \over 3}  \right)
\right.
\nn\\
&&\qquad\qquad\qquad\qquad\qquad
\left.
-{2\over 3} C_A \ln  {2\over C_F\als} +{29\over 18}C_A 
+ {2\over 3} C_F - {14\over 9}T_Fn_f \right] \Bigg)
\nn\\
&& 
+  \left[ \langle {\bf S}_{12} \rangle_{1j1} -  \langle {\bf S}_{12} \rangle_{1j'1} \right]  
\Bigg( 1+ {\als \over \pi}
\left[ \beta_0 \left( 2 \ln {2 \mu  \over m C_F \als} + {173\over 36} - {\pi^2 \over 3}  \right)
\right.
\nn\\
&&\qquad\qquad\qquad\qquad\qquad
\left.
- C_A \ln {2\over C_F\als} +{11\over 6}C_A + C_F - {4\over 3}T_Fn_f \right] \Bigg)\Bigg\}.
\label{1Phfs}
\eea
Equation (\ref{1Phfs}) is the main result presented here.

The energy splitting between the centre-of-gravity energy $E(1^3P)_{\rm c.o.g.} = 
[5\, E(1^3P_2) + 3 \, E(1^3P_1) + E(1^3P_0)]/9$ and the  $1^1P_1$ state is
entirely given at order $(C_F\als)^4\,\als/\pi$ by the spin-spin potential 
at NLO and is, therefore, not affected by our analysis.
It has been considered in \cite{Titard:1993nn,Titard:1994id,Titard:1994ry} and 
we reproduce it here for completeness: 
\be
E(1^3P)_{\rm c.o.g.} - E(1^1P_1) = { m (C_F \, \als)^4 \over 288} {\als \over \pi}
\left[ {\beta_0\over 2} - {7\over 4}C_A \right].
\ee

\section{Application to the $\chi_b(1P)$ splittings}
\label{secres}
In this section we apply Eq.~(\ref{1Phfs}) to the fine 
splittings of the bottomonium $1P$ states, $\chi_b(1P)$.
As argued in Sec.~\ref{secint} a necessary condition for this to make sense is that $p\gg\lQ$.  
In order to have an idea of the size of the scale $p$, we may follow \cite{Titard:1993nn} 
and solve the (perturbative) self-consistency equation $m C_F \als(p)/(2 n) = p$ for $n=2$. 
We obtain $p \approx 0.9$ GeV for a choice of the mass of 4.73 GeV (i.e. half of the $\Upsilon(1S)$ mass).
For comparison, in the charmonium ground state case we obtain 
$p  \approx 0.8$ GeV for a choice of the mass of 1.55 GeV (i.e. half of the $J/\psi$ mass). 
We see that it is reasonable to expect the typical momentum transfer scale for $n=2$ bottomonium states 
to be larger than the scale of non-perturbative QCD ($\siml 0.5$ GeV). 
Noteworthy it is also larger than the typical momentum transfer scale 
of the charmonium ground state, which has been often assumed to be in the perturbative
regime \cite{Titard:1993nn,Titard:1994id,Titard:1994ry,Pineda:1997hz,Brambilla:2001fw,Kniehl:2003ap}.
Indeed, $p$ has been assumed larger than $\lQ$  for  $\chi_b(1P)$ states 
in \cite{Titard:1993nn,Titard:1994id,Titard:1994ry,Pineda:1997hz,Brambilla:2001fw,Brambilla:2001qk}. 

\begin{figure}[ht]
\makebox[-9truecm]{\phantom b}
\put(20,0){\epsfxsize=7.5truecm \epsfbox{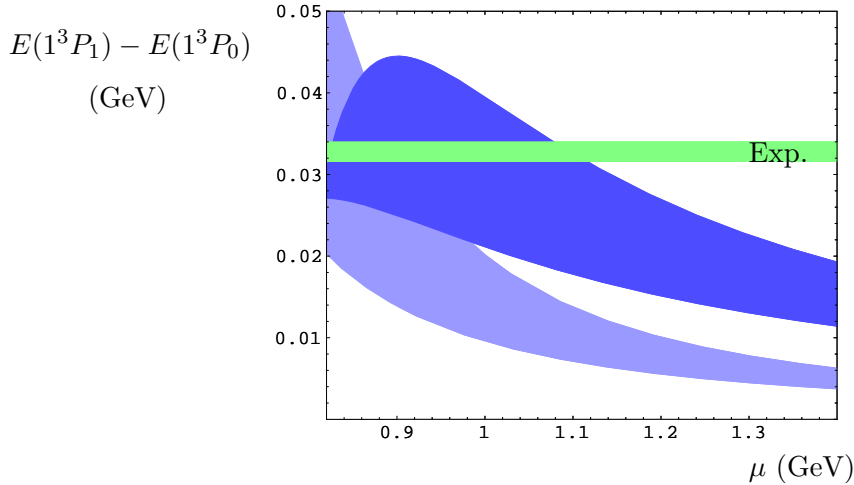}}
\put(200,-10){$\mu$ (GeV)}
\put(-80,150){$E(1^3P_1)-E(1^3P_0)$}
\put(-50,130){(GeV)}
\put(200,109){Exp.}
\caption {\it $E(1^3P_1)-E(1^3P_0)$ versus the normalization scale $\mu$. 
The light band shows the LO expectation, the dark one the NLO one as from Eq. (\ref{1Phfs}). 
The widths of the bands account for the uncertainty in $\als(M_Z) = 0.1187 \pm
0.002$ \cite{Eidelman:2004wy}. The horizontal band shows the experimental value 
$32.8 \pm 1.2$ MeV \cite{Eidelman:2004wy}. 
}
\label{figE10}
\end{figure}

\begin{figure}[ht]
\makebox[-9truecm]{\phantom b}
\put(20,0){\epsfxsize=7.5truecm \epsfbox{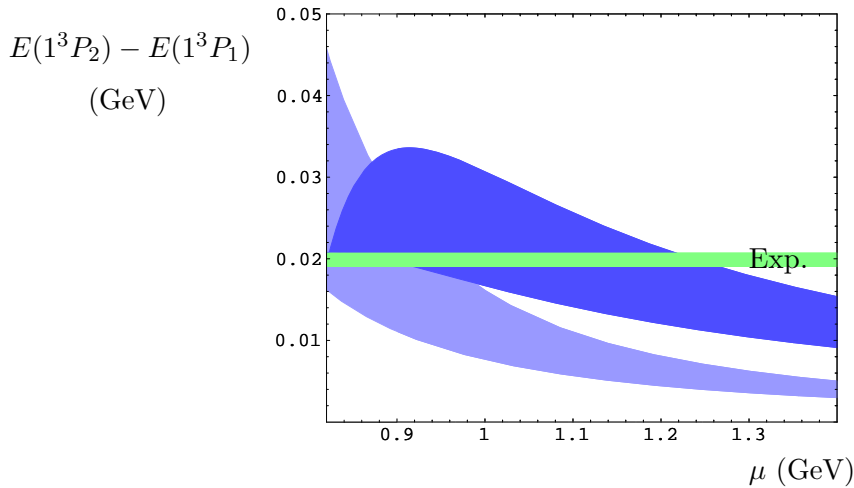}}
\put(200,-10){$\mu$ (GeV)}
\put(-80,150){$E(1^3P_2)-E(1^3P_1)$}
\put(-50,130){(GeV)}
\put(200,70){Exp.}
\caption {\it $E(1^3P_2)-E(1^3P_1)$ versus the normalization scale $\mu$. 
The horizontal band shows the experimental value $19.9 \pm 0.8$ MeV \cite{Eidelman:2004wy}.
All the rest is like in Fig.~\ref{figE10}.
}
\label{figE21}
\end{figure}

In Figs.~\ref{figE10} and \ref{figE21} we show $E(1^3P_1)-E(1^3P_0)$ and 
$E(1^3P_2)-E(1^3P_1)$ respectively as a function of $\mu$. The light and dark bands 
show the LO and NLO expectations respectively as from Eq. (\ref{1Phfs}). 
The widths of the bands account for the uncertainty in $\als(M_Z)$ only. 
The mass has been chosen to be $m=4.73$ GeV.
The correct number of massless flavours in the bottomonium case is  $n_f =3$ \cite{Brambilla:2001qk}. 
We note that in the momentum region around the physically motivated momentum transfer of about 
$0.9$ GeV the LO and NLO bands overlap with each other and with the experimental 
value. This means that the perturbative series shows convergence inside the
physically motivated momentum transfer region and reproduces, inside the uncertainties, 
the experimental values.  Note that the NLO curves have a relative maximum,
i.e. a minimal sensitivity point, that goes from 0.82 GeV to 0.90 GeV 
for $E(1^3P_1)-E(1^3P_0)$ and from 0.83 GeV to 0.91 GeV for
$E(1^3P_2)-E(1^3P_1)$ as $\als(M_Z)$ goes from 0.1167 to 0.1207.
Therefore, the physically motivated momentum transfer region overlaps  
also with the region covered by the minimal sensitivity points of the NLO curves.

\begin{figure}[htb]
\makebox[-9truecm]{\phantom b}
\put(20,0){\epsfxsize=7.5truecm \epsfbox{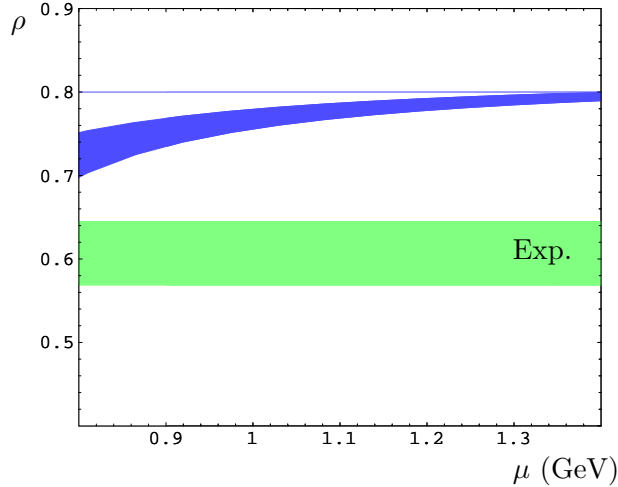}}
\put(200,-10){$\mu$ (GeV)}
\put(10,160){$\rho$}
\put(200,73){Exp.}
\caption {\it $\rho$ versus the normalization scale $\mu$. The horizontal line at 0.8 corresponds 
to the LO expectation, the  curve to the NLO one. The band accounts for the uncertainty in $\als(M_Z) = 
0.1187 \pm 0.002$ \cite{Eidelman:2004wy}. The horizontal band at $0.607 \pm 0.038$ 
shows the experimental value \cite{Eidelman:2004wy}.
}
\label{figrho}
\end{figure}

The agreement between the fine splittings calculated at NLO and the
experimental data appears reasonably good. Moreover it happens in a region where
the physical momentum transfer is expected to be, the perturbative series shows convergence 
and the NLO curve is less dependent on the scale $\mu$. However, the uncertainties 
are large so that no conclusive statement can be made at this point.
A quantity less sensitive to $\als$ and to correlated uncertainties 
in the splittings is the ratio 
\be
\rho = {E(1^3P_2) - E(1^3P_1) \over E(1^3P_1) - E(1^3P_0)}.
\label{rho}
\ee
It has been suggested long ago that this observable may be rather sensitive 
to non-perturbative contributions in the spin-dependent potentials
\cite{Schnitzer:1978kf}. The ratio $\rho$ is equal to 0.8 at LO in
perturbation theory, while it is measured to be $0.607 \pm 0.038$ 
for the $1P$ bottomonium state \cite{Eidelman:2004wy}.
The dark band in Fig.~\ref{figrho} shows the NLO expectation.
We see that the NLO corrections go toward the data and may explain from about 
$15\%$ to $65\%$ of the difference between them and the LO value in the 
scale range $0.8 \div 1$ GeV. 
At the scale $\mu = 0.9$ GeV the NLO calculation seems to account for $20\% \div 40\%$ of the 
difference between 0.8 and the experimental value. 
At this stage we cannot say to what extent the remaining difference is due to non-perturbative
corrections, since higher-order perturbative corrections not included 
in this analysis can still be large, as the scale dependence of $\rho$ and the case 
of the $n=1$ hyperfine splitting may indicate.
In this respect it would be important to perform a 
NLL analysis similar to those done in \cite{Kniehl:2003ap,Penin:2004xi}.

\bigskip

{\bf Acknowledgements.}
We thank Francisco J. Yndurain for communications.

\end{document}